\documentclass[%
reprint,
amsmath,amssymb,
aps,
prb,
floatfix,
]{revtex4-2}

\UseRawInputEncoding
\usepackage{graphicx}% Include figure files
\usepackage{dcolumn}% Align table columns on decimal point
\usepackage{bm}% bold math
\usepackage{txfonts}
\usepackage{float}
\usepackage{soul, color, xcolor}
\let\hl\relax
\soulregister\cite7
\soulregister\ref7 
\begin{document}
%\begin{CJK}{UTF8}{gbsn}

\preprint{APS/123-QED}

\title{Local optical conductivity of strain solitons in bilayer graphene with arbitrary soliton angle}% Force line breaks with \\

\author{Lu Wen}
\author{Xinyu Lv}%
\author{Zhiqiang Li}%
 \email{zhiqiangli@scu.edu.cn}
\affiliation{%
 College of Physics, Sichuan University, Chengdu, Sichuan 610064, China}%

\date{\today}% It is always \today, today,
             %  but any date may be explicitly specified

\begin{abstract}
We theoretically study the electronic band structure and local optical conductivity of domain wall solitons in bilayer graphene (as well as twisted bilayer graphene) with arbitrary soliton angle, which characterizes the local strain direction. We demonstrate that the soliton angle provides an important yet underexplored degree of freedom that can strongly modify the local optical conductivity. The conductivity spectrum features resonance peaks associated with interband transitions involving the topological as well as high-energy soliton states. Two most prominent peaks exhibit continuous suppression and enhancement, respectively, with the soliton angle. The dependence of the peaks on Fermi energy provides important information about the soliton band structure. The local optical conductivity exhibits substantial spatial dependence, which can be used to study the spatial distribution of the soliton states. Furthermore, we show that the conductivity spectra for all soliton angles are broadly tunable by external pressure, which can double the energies of the resonance peaks in experimentally achievable pressure ranges. 
\end{abstract}

\maketitle

%\tableofcontents

\section{\label{sec:level1} INTRODUCTION}
Twisted bilayer graphene (TBG) formed by stacking two graphene layers with a twist angle can host a wide variety of quantum effects and phases, such as superconductivity\cite{RN127,RN389} and correlated insulating phases\cite{RN126} near the ‘magic’ angle $\theta \approx  1.1^\circ$. Moreover, minimally twisted bilayer graphene (mTBG) ($\theta \ll 1^\circ$)\cite{RN279} provides a fertile ground to explore novel quantum phenomena. Four different stacking configurations can be found in TBG: AA, AB, BA, and saddle point (SP). In mTBG, the energetically favourable Bernal stacked regions (AB and BA) are maximized at the expense of AA sites, due to lattice relaxation and reconstruction\cite{RN279,RN364}. As a result, the Bernal stacked domains are separated by a triangular network of solitons (i.e. domain walls) with concentrated strain\cite{RN395}. Recently, a plethora of new electronic states and phenomena have been reported in such soliton networks, for instance, chiral one-dimensional (1D) topological states\cite{RN495,RN501,RN190}, 1D proximity superconductivity\cite{RN587}, deeply confined electronic states\cite{RN494} and various exotic electronic phases\cite{RN496}. In addition, recent theoretical studies have predicted many novel phases in soliton networks in mTBG, including Floquet topological insulators\cite{RN499,RN500}, charge density wave\cite{RN498}, Luttinger liquid phase\cite{RN497} and other correlated state\cite{RN497}.

The soliton networks also provide a promising platform to explore many unique optical phenomena in mTBG, such as photonic crystals for plasmons\cite{RN132}, plasmonic Dirac cone\cite{RN524} and nanoscale directional photocurrent\cite{RN421,RN392}. As for polaritonic response, polaritons can be strongly scattered by the solitons due to the local optical conductivity enhancement at solitons with respect to AB (BA) domain. Soliton networks act as the quantum photonic crystal with a periodic array of scatterers, which can thus control the polariton propagation in soliton superlattices\cite{RN132}. Furthermore, local optical conductivity in soliton network is directly related to the light absorption, which significantly affects the distinctive optoelectronic response at solitons via photo-thermoelectric effect\cite{RN421,RN392}. Therefore, a new approach for controlling local optical conductivity at solitons can provide new opportunities for exploring the polaritonic physics and optoelectronic response in mTBG and related two-dimensional (2D) moir\'{e} structures.

\begin{figure*}
\includegraphics[width=2\columnwidth]{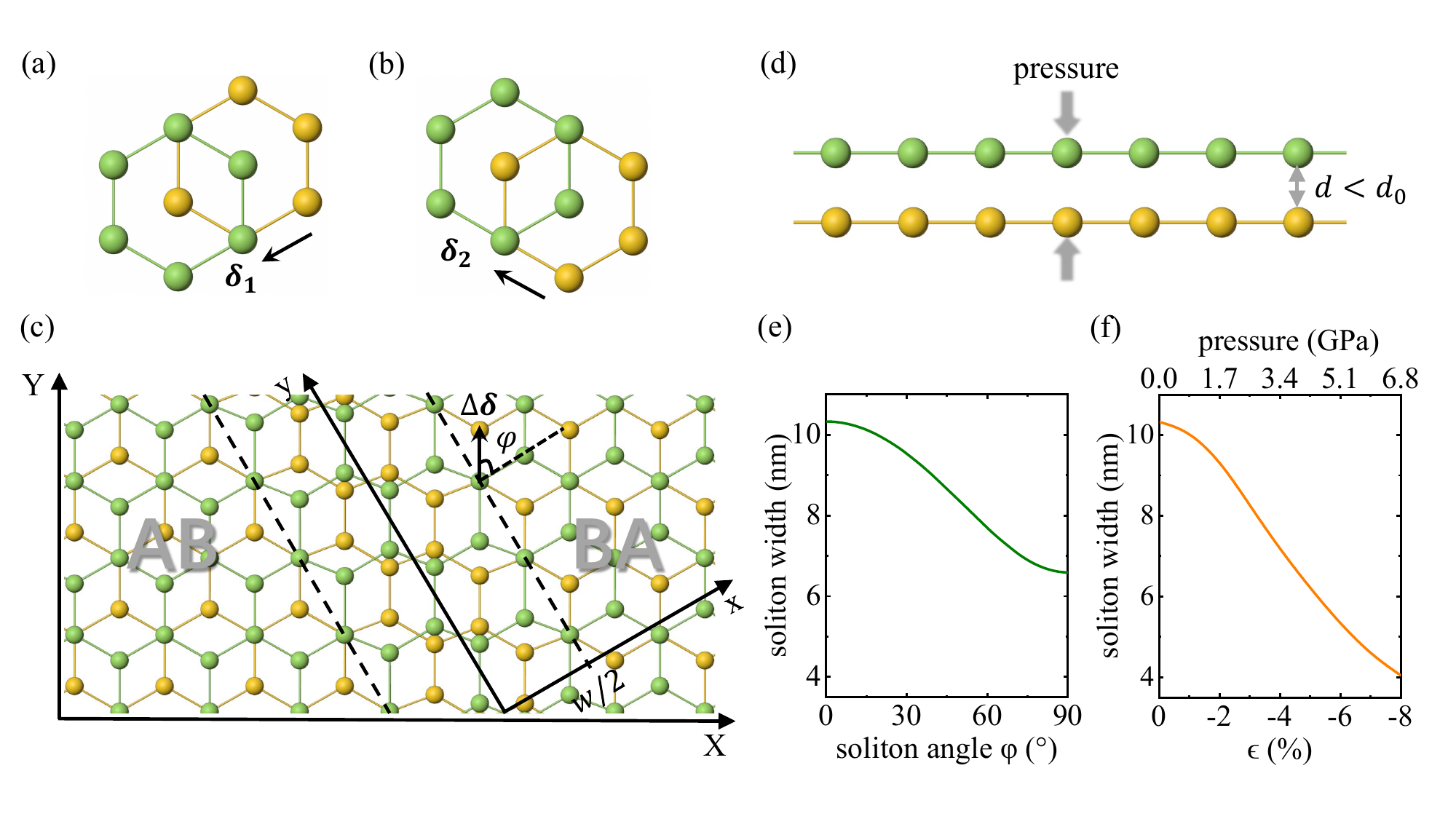}% Here is how to import EPS art
\caption{\label{fig:structure}(Color online) The in-plane displacement vector $\boldsymbol{\delta}_1$ and $\boldsymbol{\delta}_2$ between the two graphene sheets in uniform (a) $\mathrm{AB}$ and (b) $\mathrm{BA}$ domain. Green and orange hexagons represent the lattices of layers 1 and 2, respectively. (c) Schematic structure of soliton with arbitrary soliton angle $\varphi$. The black dashed lines represent boundaries of the soliton in the middle. The angle between the local change $\Delta \boldsymbol{\delta}$ in $\boldsymbol{\delta}$ and the soliton normal is defined as the soliton angle $\varphi$. (d) Schematic diagram of external pressure applied to bilayer graphene, reducing the interlayer spacing $d$ between adjacent layers. The dependence of soliton width $w$ on (e) soliton angle $\varphi$ and (f) pressure $\boldsymbol{\epsilon}$ are presented, respectively. In ( $\mathrm{f}$ ), we take the tensile soliton $\left(\varphi=0^{\circ}\right)$ as an example.}
\end{figure*}

The soliton angle is a critical parameter for characterizing the local strain direction at solitons. In mTBG, there is an in-plane displacement vector $\boldsymbol{\delta}_1\left(\boldsymbol{\delta}_2\right)$ between the two graphene sheets in AB (BA) Bernal stacked domain, as shown in Fig.~\ref{fig:structure}(a) and (b). When traversing different domains, a local change $\Delta \boldsymbol{\delta}$ in interlayer displacement $\boldsymbol{\delta}$ is inevitably required at the soliton with saddle-point (SP) stacking in the middle. The physical properties of solitons are governed by the local soliton angle $\varphi-$ the angle between the local strain direction $\Delta \boldsymbol{\delta}$ and the soliton normal\cite{RN395} as shown in Fig.~\ref{fig:structure}(c). Two limiting cases are $\varphi = 90^\circ$ for shear solitons and $\varphi = 0^\circ$ for tensile solitons. Practically, all intermediate configurations are also possible. The inhomogeneities in strain and twist angle are ubiquitous in realistic mTBG samples\cite{RN502}, leading to distorted inhomogeneous triangular soliton networks and thus spatial variations in soliton angles\cite{RN494,RN496,RN437,RN512}. 

\hl{The critical role of soliton angle in governing the physical properties of solitons and soliton superlattices in bilayer graphene  has been demonstrated in earlier studies. Initially, the continuous soliton-angle variations within the soliton networks were investigated, revealing the importance of soliton angle in determining the soliton width and energy of the soliton\cite{RN395}. Subsequently, although the effect of various soliton angle on nanoscale optoelectronic response across solitons was experimentally observed\cite{RN421}, the significance of soliton angle was neglected at the time. Recently, advances in fabrication techniques have made it possible to create and control solitons with arbitrary soliton angles\cite{RN494,RN592,RN505,RN486}, which is crucial for further experimental exploration. Moreover, combined theoretical and experimental study of electron transport across solitons with varying soliton angle showed the transport gap that can be significantly tuned through soliton angle in bilayer graphene\cite{RN443}. M. Koshino pointed out that electron transport across shear and tensile solitons strongly depends on the soliton direction and also on the atomic configuration inside\cite{RN331}. All these previous works have highlighted that the soliton angle is a very important degree of freedom in 2D systems.}

However, the optical properties of solitons with arbitrary soliton angles other than shear and tensile walls have remained largely unexplored to date. Previously, the local optical conductivities for shear and tensile solitons in bilayer graphene were calculated for selected single frequency\cite{RN132,RN439}. More recently, the local optical conductivity spectra for shear solitons in bilayer graphene were calculated over a range of frequencies\cite{RN306}. Note that the variation of soliton angle can strongly modify the local band structure and optical properties of the soliton \cite{RN439,RN55,RN443}. Since the variations of local physical properties of 1D solitons can crucially affect the global properties of the entire soliton network and lead to new physics in mTBG, it is of great importance to study the local optical conductivity of solitons with arbitrary soliton angles.

In this article, we perform systematic investigations on the band structure and local optical conductivity of 1D solitons with arbitrary soliton angle in bilayer graphene, which is directly relevant to the optical properties of mTBG and related 2D moir\'{e} structures. We study the conductivity spectrum in the mid-infrared region with the Fermi energy above the bulk bandgap, which is the parameter regime similar to those in recent experiments studying nanophotonic effects in mTBG \cite{RN132,RN439,RN306,RN55}. We find that the soliton angle can drastically alter the soliton band structure, including both the topological states and high-energy soliton states. The interband transitions involving these states result in a series of distinctive resonance peaks in the local optical conductivity. Two most prominent peaks are continuously suppressed or enhanced, respectively, by increasing the soliton angle, and their sensitive dependence on doping can be used to identify important features in the soliton band structure. Next, we discuss the spatial distribution of local optical conductivity across a soliton. Moreover, we show that the band structures and local optical conductivity for all soliton angles are broadly tunable by pressure applied perpendicular to the layers. In particular, the energies of resonance peaks in the conductivity spectra double at experimentally achievable pressures.

\section{THEORETICAL METHODS}

In our calculations, we consider a single infinitely long domain wall soliton with an arbitrary soliton angle $\varphi$ in bilayer graphene, starting with a uniform and infinite AB-stacked domain to define the atomic structure as shown in Fig.~\ref{fig:structure}(c). We set the $x$ and $y$-axis perpendicular and parallel to the soliton, as well as the $X$ - and $Y$-axis along the zigzag and armchair directions of the graphene lattice, respectively. The entire system is divided into three components: the AB $(x<-w / 2)$, soliton $(-w / 2<x<w / 2)$ and BA region $(x>w / 2)$, where $w$ represents the width of the soliton. In the $X-Y$ coordinate system, the primitive lattice vectors of graphene are $\boldsymbol{a}_{\boldsymbol{1}}=\sqrt{3} a_0 \boldsymbol{e}_{\boldsymbol{X}}$ and $\boldsymbol{a}_{\mathbf{2}}=(\sqrt{3} / 2) a_0 \boldsymbol{e}_{\boldsymbol{X}}-(3 / 2) a_0 \boldsymbol{e}_{\boldsymbol{Y}}$, where $\boldsymbol{e}_{\boldsymbol{X}}$ and $\boldsymbol{e}_{\boldsymbol{Y}}$ are respectively the unit vectors along $X-$ and $Y-$ axis, and $a_0 \approx 0.142 \mathrm{~nm}$ is the length of carbon-carbon bond in graphene.

Based on the continuum model for bilayer graphene with a spatial dependent interlayer displacement vector $\boldsymbol{\delta}(x)=\delta_X(x) \boldsymbol{e}_{\boldsymbol{X}}+\delta_Y(x) \boldsymbol{e}_{\boldsymbol{Y}}$, the low energy effective Hamiltonian near the $K\left(K^{\prime}\right)$ valley in the basis of sublattices of layers 1 and $2\left(A_1, B_1\right.$, $\left.A_2, B_2\right)$ is written as\cite{RN439,RN443,RN331} 
\begin{equation}
H=\left[\begin{array}{cc}
H_0^{+} & U^{\dagger}(x) \\
U(x) & H_0^{-}
\end{array}\right]
\label{eq:one}.
\end{equation}
Here, $\mathit{H}_0$ is the Dirac Hamiltonian for a monolayer graphene given by\cite{RN443} 
\begin{equation}
H_0^{ \pm}=\left[\begin{array}{cc} 
\pm \frac{V}{2} & \hbar v\left(\xi k_x+i k_y\right) e^{i \phi} \\
\hbar v\left(\xi k_x-i k_y\right) e^{-i \phi} & \pm \frac{V}{2}
\end{array}\right]
\label{eq:two},
\end{equation}
where $V=e V_i$ is the interlayer potential due to an external bias $V_i$, $\xi= \pm 1$ labels the valley index corresponding to $K$ and $K^{\prime}$, $e^{-i \phi}$ is introduced to account for soliton angle $\varphi$, and $\phi=\pi / 2-\varphi$ denotes the angle between the armchair direction of the graphene lattice and the soliton ($y-$axis). 

The $U(x)$ term in Eq.~(\ref{eq:one}) describes the interlayer interaction in bilayer graphene shifted by a spatially varying displacement $\boldsymbol{\delta}(x)$, which can be expressed as
\begin{equation}
U(x)=\left[\begin{array}{ll}
U_{A_2 A_1} & U_{A_2 B_1} \\
U_{B_2 A_1} & U_{B_2 B_1}
\end{array}\right]=\left[\begin{array}{cc}
u\left(\boldsymbol{K}_{\xi}, \boldsymbol{\delta}\right) & u\left(\boldsymbol{K}_{\xi}, \boldsymbol{\delta}+\boldsymbol{\tau}\right) \\
u\left(\boldsymbol{K}_{\xi}, \boldsymbol{\delta}-\boldsymbol{\tau}\right) & u\left(\boldsymbol{K}_{\xi}, \boldsymbol{\delta}\right)
\end{array}\right]
\label{eq:three}.
\end{equation}
In Eq.~(\ref{eq:three}), $\boldsymbol{\tau}$ is the vector from the B sites to the nearest A sites in graphene lattice. The function $u\left(\boldsymbol{K}_{\xi}, \boldsymbol{\delta}\right)$ is approximately written in terms of a few Fourier components as $u\left(\boldsymbol{K}_{\xi}, \boldsymbol{\delta}\right)=\frac{\gamma_1}{3}\left(1+e^{-i \xi \boldsymbol{a}_1^* \boldsymbol{\delta}}+e^{-i \xi\left(\boldsymbol{a}_1^*+\boldsymbol{a}_2^*\right) \boldsymbol{\delta}}\right)$\cite{RN331}, where $\boldsymbol{a}_i^*$ is the reciprocal lattice vector of graphene satisfying $\boldsymbol{a}_{\boldsymbol{i}} \cdot \boldsymbol{a}_j^*=2 \pi \delta_{i j}$ and $\gamma_1$ is the interlayer coupling strength. Trigonal warping terms $\gamma_3$ and $\gamma_4$ have been neglected for simplicity. 

To describe the local properties at solitons in bilayer graphene, the homogeneous Hamiltonian $H$ has to be modified to take into account the spatial variation of displacement $\boldsymbol{\delta}(x)$. In the $x-y$ coordinate system, the momentum $k_y$ remains a good quantum number due to the translation invariance. The momentum $k_x$ perpendicular to the soliton is replaced by the operator $-i \partial / \partial x$. \hl{In the numerical calculation, the total system is discretized in the $x$ direction into $N_x+1$ grid points $(x_0,x_1,...,x_i,...,x_{N_x-1},x_{N_x})$ with $\Delta {x}=x_{N_x}-x_{N_x-1}$.} Consequently, the real space Hamiltonian $H\left(x, k_y\right)$ is solved numerically on a 1D grid along $x$ direction \hl{using the finite differences method. Typically, $N_x=1200$ and $\Delta {x}=0.25\mathrm{~nm}$.} As shown in Fig.~\ref{fig:structure}(c), the AB and BA domains on either side of the soliton respectively have stacking vectors $\boldsymbol{\delta}_1=-(\sqrt{3} / 2) a_0 \boldsymbol{e}_{\boldsymbol{X}}-(1 / 2) a_0 \boldsymbol{e}_{\boldsymbol{Y}}$ and $\boldsymbol{\delta}_{\boldsymbol{2}}=-(\sqrt{3} / 2) a_0 \boldsymbol{e}_{\boldsymbol{X}}+(1 / 2) a_0 \boldsymbol{e}_{\boldsymbol{Y}}$ in the $X-Y$ frames. Thus, the interlayer translation vector $\Delta \boldsymbol{\delta}=\boldsymbol{\delta}_2-\boldsymbol{\delta}_1$ associated with the soliton must be along the $Y$(armchair) direction. Following the previous transmission electron microscopy studies, the spatial distribution of $\boldsymbol{\delta}(x)$ across a soliton has the form\cite{RN395,RN507} 
\begin{equation}
\boldsymbol{\delta}(x)=\frac{2}{\pi} \arctan \left(e^{\pi x / w}\right) a_0 \boldsymbol{e}_{\boldsymbol{Y}}+\boldsymbol{\delta}_1
\label{eq:four}.
\end{equation}
The stacking order $\delta \in[1,2]$ with $\delta=1,1.5,2$ corresponding to AB, SP, and BA stacking in a bilayer system. 

Owing to competition between strain energy in the soliton and the misalignment energy cost per unit length of the soliton, the widths $w$ of the solitons are different in all soliton configurations, which are given by the two-chain Frenkel–Kontorova model\cite{RN395}:
\begin{equation}
w=\frac{a_0}{2} \sqrt{\frac{1}{V_{s p}}\left(\frac{E t}{1-\nu^2} \cos ^2 \varphi+G t \sin ^2 \varphi\right)}
\label{eq:five}.
\end{equation}
Here, $Et=340 \mathrm{~N} / \mathrm{m}$ is Young's modulus, $Gt\sim Et /(2(1+\nu))=142 \mathrm{~N} / \mathrm{m}$ is the shear modulus with the Poisson ratio $\nu=0.16$, and $V_{s p}$ represents saddle-point energy per unit area. In bilayer graphene, the interlayer spacing is a critical variable for understanding the elastic distortion of the bilayer. Accordingly, $V_{s p}$ as a function of interlayer spacing $d$ can be obtained from the fitting results of three-dimensional generalized stacking-fault energy (GSFE) predicted by theory\cite{RN460}, which describes the difference of energy between the ground-state structure and the uniform nonaligned structure. $V_{s p}=1.2 \mathrm{meV/atom}$ has a minimum value in the absence of external disturbance\cite{RN395}.  

Generally, the interlayer spacing $d$ of bilayer graphene is particularly responsive to some external perturbations, such as the pressure along the direction perpendicular to the layers\cite{RN513,RN515} in Fig.~\ref{fig:structure}(d). For experimentally accessible perturbation, the vertical compression of the bilayer is defined in terms of $\epsilon=\left(d / d_0\right)-1$, which corresponds to an external pressure of $P=A\left(e^{-\epsilon B}-1\right)$ with $A=5.73 \mathrm{GPa}$ and $B=9.54$\cite{RN459}. Here $d_0=3.35 \AA$ is the interlayer distance without external perturbation. Below, we use the compression parameter $\epsilon$ to represent the magnitude of the external pressure. In Fig.~\ref{fig:structure}(e) and ~\ref{fig:structure}(f), we give the dependence of soliton width $w$ on soliton angle $\varphi$ and pressure $\epsilon$, respectively. Both can effectively narrow solitons, but the latter has a more dramatic effect by amplifying lattice relaxation. Without external pressure, the width of solitons with arbitrary soliton angle $\varphi$ is about 6.2 to $10.1\mathrm{~nm}$\cite{RN395} as seen in Fig.~\ref{fig:structure}(e). In addition, the reduction of parameter $d$ under pressure $\epsilon$ can change the distance-dependent interlayer coupling strength $\gamma_1$\cite{RN513,RN459}. Then, we focus on the relative scaling of inter- and intra-layer energies, and the interlayer coupling $\gamma_1$ has at most quadratic dependence on external pressure $\epsilon$, $\gamma_1(\epsilon)=t_2 \epsilon^2-t_1 \epsilon+t_0$ with the coefficients $t_{[0,1,2]}=[0.400,2.234,9.190] \mathrm{eV}$\cite{RN459}.

\begin{figure*}
\includegraphics[width=2\columnwidth]{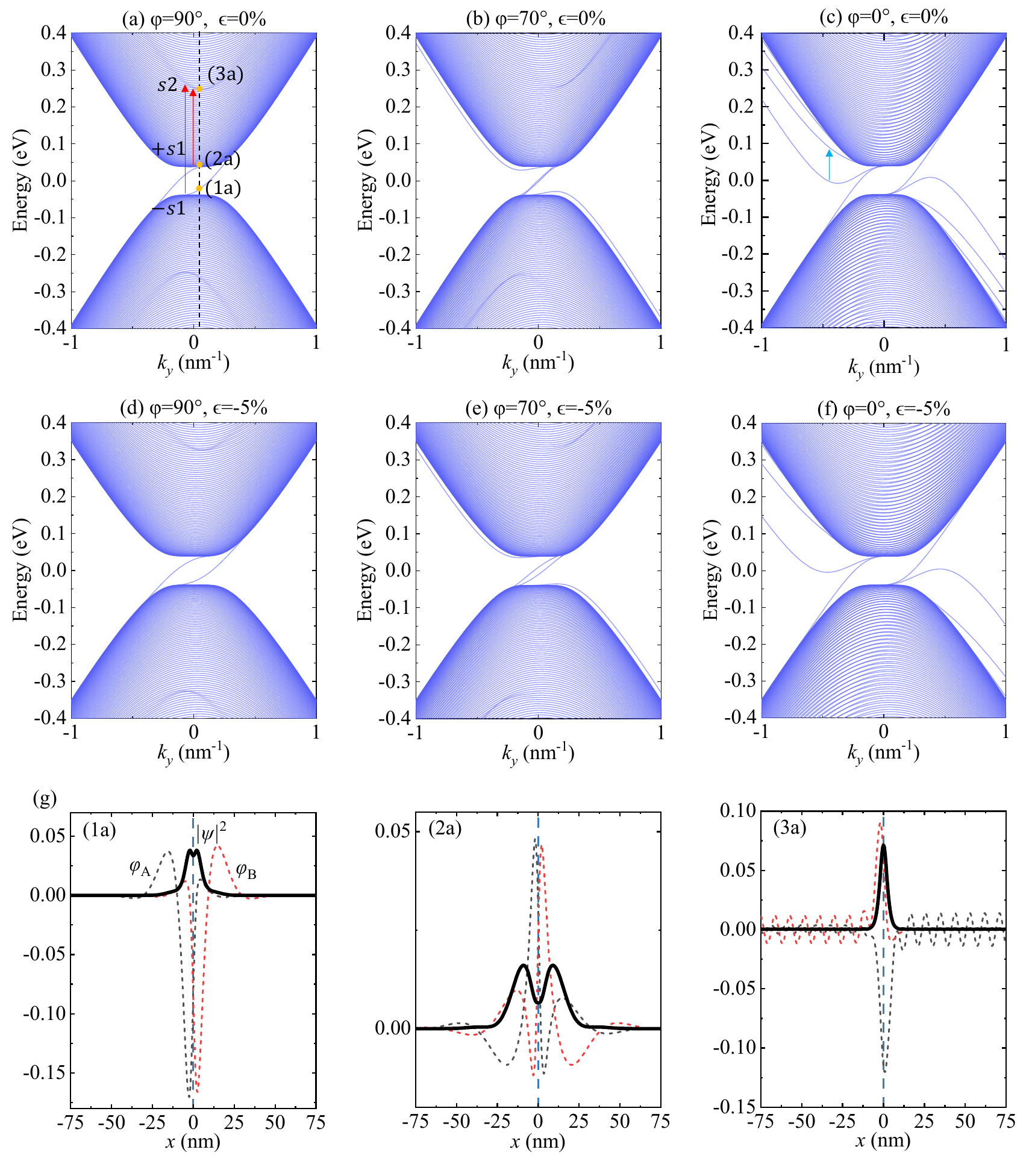}% Here is how to import EPS art
\caption{\label{fig:band}(Color online) \hl{Band structure, wave functions, and probability densities for $K$ valley.} (a)-(f) Electronic band structure of a single soliton for representative soliton angles $\varphi=90^{\circ}, 70^{\circ}, 0^{\circ}$ in bilayer graphene with no pressure in (a)-(c) and pressure $\epsilon=$ $-5 \%(P=3.6 G P a)$ in (d)-(f). \hl{(g) The real parts of the wave functions and probability densities of the topological state (plot (1a)), state $s1$ (plot (2a) and state $s2$ (plot (3a), corresponding to the three yellow points with $k_y=0.04 \mathrm{~nm}^{-1}$ indicated by the black dashed line in (a). The blue dashed lines indicate the center of soliton at $x=0\mathrm{~nm}$.} Parameter used for the calculations is $V_i=80 \mathrm{mV}$.}
\end{figure*}

Using the eigenvalues $\left(E_m, E_n\right)$ and eigenfunctions $\left(\left|a_m\right\rangle,\left|a_n\right\rangle\right)$ obtained from the above continuum model, the nonlocal optical conductivity $\Sigma\left(x, x^{\prime}\right)$ of bilayer graphene can be calculated by the Kubo formula \cite{RN439}
\begin{equation}
\begin{aligned}
\Sigma_{\alpha \alpha}\left(x, x^{\prime}\right)= & \frac{g_s g_v e^2 \hbar}{4 \pi^2 i} \int d k_y\left(\sum_{m \neq n} \frac{f_m-f_n}{\left(E_m-E_n\right)\left(\hbar \omega+i \eta-\left(E_m-E_n\right)\right)}\right. \\
& \left.+\sum_n \frac{d f_n}{d E_n} \frac{1}{\hbar \omega+i \eta}\right) M_\alpha^*(x) M_\alpha\left(x^{\prime}\right) .
\end{aligned}
\label{eq:six}
\end{equation}
Here $\alpha=x$ or $y, x$ and $x^{\prime}$ are coordinates in the direction normal to the soliton, $g_s=$ $g_v=2$ denotes the spin and valley degeneracy in graphene, $f_m\left(f_n\right)$ is the Fermi-Dirac distribution, $\eta$ is the phenomenological damping rate, and $\hbar \omega$ is the incident photon energy. The matrix element is defined as $M_\alpha=\left\langle a_m\left|v_\alpha\right| a_n\right\rangle$, and $v_\alpha=\frac{\partial H}{\hbar \partial k_\alpha}$ represents the velocity operator. Then the local optical conductivity can be calculated from $\sigma(x) \equiv \int \Sigma\left(x, x^{\prime}\right) d x^{\prime}$\hl{, where $\Sigma\left(x, x^{\prime}\right)$ is a matrix of $(N_x+1)*(N_x+1)$}.

\section{RESULTS AND DISCUSSION}
The electronic band structures of the solitons with various soliton angle $\varphi$ and pressure index $\epsilon$ are plotted in Fig.~\ref{fig:band} for the positive interlayer bias $V_i$. \hl{These diagrams can be qualitatively understood as the superposition of local band structures in the soliton region ($-w / 2<x<w / 2$) along with AB (BA) domain ($x<-w/2$ and $x>w/2$)} projected from the $2 \mathrm{D}$ momentum space to the $1 \mathrm{D}$ momentum axis $k_{\mathrm{y}}$ parallel to the domain wall. In the cases of shear $\left(\varphi=90^{\circ}\right)$ and tensile $\left(\varphi=0^{\circ}\right)$ walls without external pressure in Fig.~\ref{fig:band}(a) and ~\ref{fig:band}(c), we obtain similar results to those reported in previous studies\cite{RN439}. For all cases shown in Fig.~\ref{fig:band}, a finite interlayer bias $V_i$ can induce a bandgap in the band continua of bulk $\mathrm{AB}$ (BA) bilayer graphene. However, there is still a pair of topological chiral states connecting valence and conduction bands, because the valley Chern numbers change sign from $\mathrm{AB}$ to $\mathrm{BA}$ region across the soliton\hl{\cite{RN449,PhysRevX.3.021018,RN439,RN132}}. Strikingly, the energy separation between the topological states can be effectively controlled by the soliton angle $\varphi$, which decreases and then increases from the shear to tensile soliton. Except for the topological states, conventional (non-topological) bound states separated from the band continua also appear in non-shear solitons, the number of which is directly linked to the soliton width $w$ in virtue of the different quantum confinement effects\cite{RN439}. 
\hl{The pressure enhances the interlayer interactions between two layers\cite{RN459}, therefore increasing the saddle-point stacking energy $V_{sp}$\cite{RN460}. From Eq.~(\ref{eq:five}), the sharper domain walls are generated by pressure in bilayer graphene because of the larger $V_{sp}$. Hence, more conventional bound states are pushed towards the band continua due to the stronger quantum confinement at the soliton, thereby making some state branches indistinguishable from the boundary states under pressure, as shown in Fig.~\ref{fig:band}(c) and ~\ref{fig:band}(f).}

\hl{Interestingly, we also find dispersing high-energy states $s1$ close to bulk band boundaries and $s2$ near $E=0.25 \mathrm{eV}$ inside the bulk band continua as shown in Fig.~\ref{fig:band}(a). To investigate the physical origin of these states, we plot the real part of the wave functions of sublattice A (black dotted curves) and B (red dotted curves) as well as the probability densities for the topological states, high-energy states $s1$ and $s2$ in Fig.~\ref{fig:band}(g), plots (1a)-(3a), respectively, indicated by the yellow dots corresponding to $k_y=0.04 \mathrm{~nm}^{-1}$ in Fig. ~\ref{fig:band}(a). Regarding chiral topological states, we obtain results in plot (1a) that are consistent  with those in the previous literature\cite{RN448,RN510}. The electronic states $s1$ and $s2$ are also bound in the $x$ direction near the soliton at $x=0 \mathrm{~nm}$ as seen in plots (2a) and (3a). However, the wave functions of bound states $s1$ are more extended than the topological states and $s2$, and have a clear nodal character near the center of soliton, which are similar to the reported results\cite{RN510}. Moreover, the probability distribution, $|\psi|^2=\left|\psi_A\right|^2+\left|\psi_B\right|^2$, of high-energy states $s1$ and $s2$ is slightly different from that of chiral topological states. Compared to the topological states, the states s1 are less localized while $s2$ are strongly confined in the soliton region.}

\begin{figure*}
\includegraphics[width=2\columnwidth]{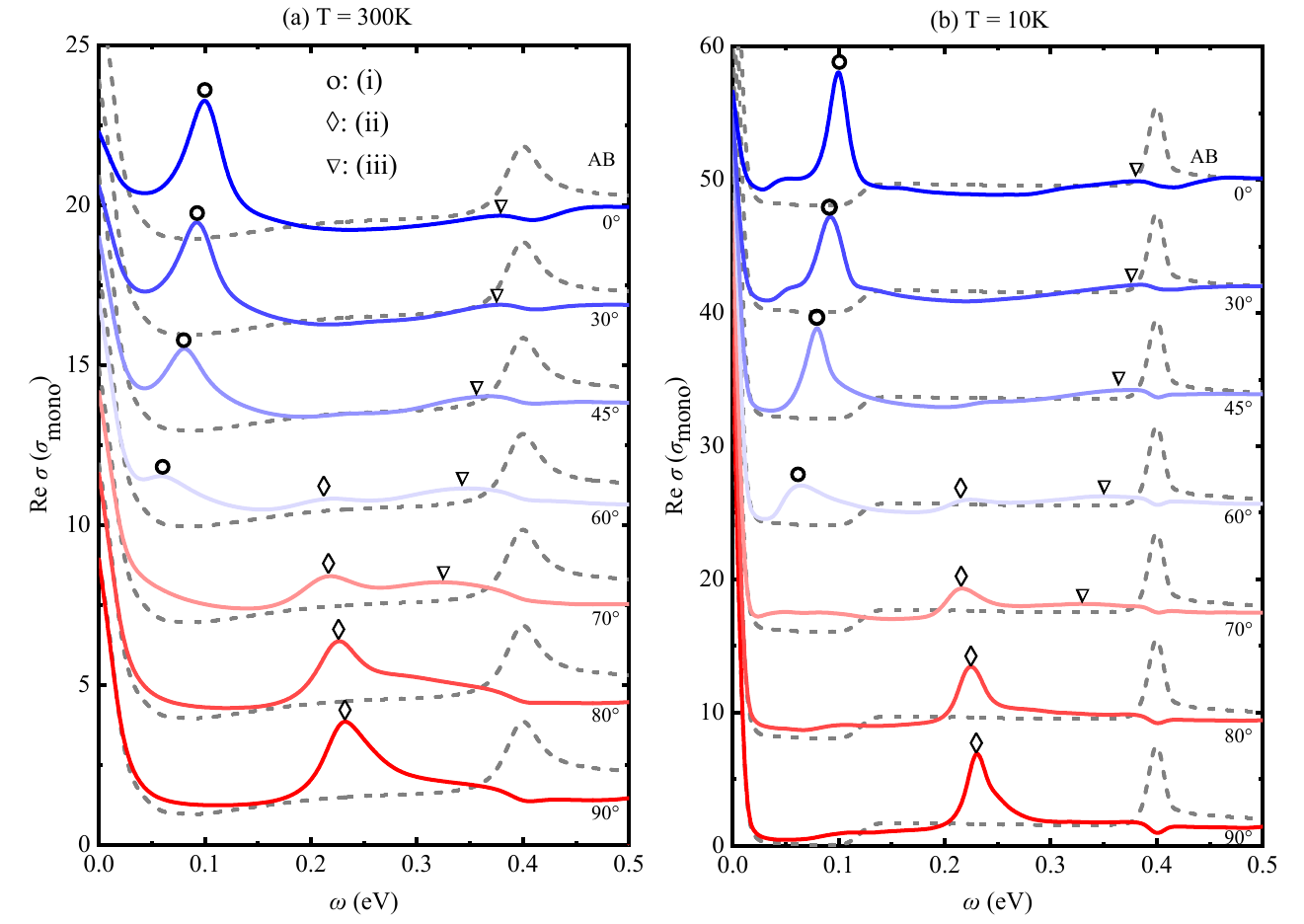}% Here is how to import EPS art
\caption{\label{fig:LOP1} \hl{(Color online) Real parts of the local optical conductivity spectra $\sigma(\omega)$ for the narrow band gap at different temperatures: (a) $T=300 \mathrm{~K}$, $\eta=15 \mathrm{~meV}$ and (b) $T=10 \mathrm{~K}$, $\eta=3 \mathrm{~meV}$. The gray dashed lines represent the $\sigma(\omega)$ spectra for the $AB$ domain at $x=75\mathrm{~nm}$, and the colored lines show the $\sigma(\omega)$ spectra for solitons at $x=0\mathrm{~nm}$ with $\varphi=0^{\circ}, 30^{\circ}, 45^{\circ}, 60^{\circ}, 70^{\circ}, 80^{\circ}, 90^{\circ}$. In (a) and (b), the spectra are vertically offset by 
$3\sigma_{\text{mono}}$ and $6\sigma_{\text{mono}}$ from one another for clarity, respectively. Parameters used for the calculations are $\epsilon=0 \%$, $V_i=20 \mathrm{mV}$ and $\varepsilon_F=60 \mathrm{meV}$.}}
\end{figure*}

From a close examination of the spatial profile of the wavefunction, the high-energy soliton states $s 1$ and $s 2$ are qualitatively inherited from the local band structure in the centre of soliton with SP stacking. For the narrow domain walls (width $\sim 10 \mathrm{~nm}$ ), other high-energy dispersing soliton states are pushed to the boundaries of the bulk bands due to quantum confinement effects, so only the states labeled $s 1$ and $s 2$ are visible as darker colored dispersing branches in Fig.~\ref{fig:band}. Intriguingly, the high-energy $s 2$ states are significantly sensitive to soliton angle $\varphi$ and pressure index $\epsilon$. In Fig.~\ref{fig:band}(a)-(c), the soliton band structure evolves systematically as the direction of the wall deviates from the shear soliton. Such an evolution originates from the fact that the local 2D band structures inside the different types of solitons are projected to $k_{\mathrm{y}}$ axis with the variable projection angles $\varphi$. Moreover, under the effect of external pressure $\epsilon$, the interlayer interaction between adjacent graphene layers increases because of the decreasing interlayer spacing $d$, resulting in the soliton states $s 2$ shifting to higher energies, about $E=0.32 \mathrm{eV}$ at $\epsilon=-5 \%(P=3.6 G P a)$ as illustrated in Fig.~\ref{fig:band}(d)-(f).

\begin{figure}
\includegraphics[width=1\columnwidth]{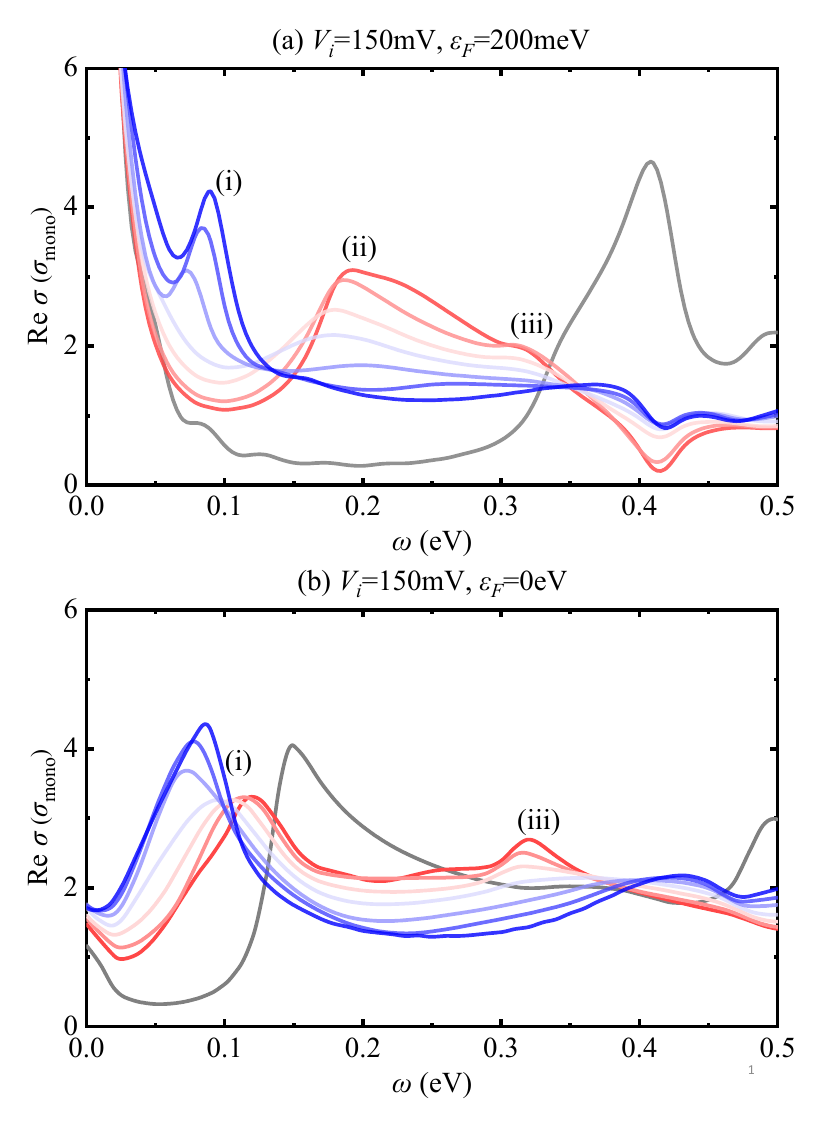}% Here is how to import EPS art
\caption{\label{fig:LOP2}(Color online) Real parts of the local optical conductivity spectra $\sigma(\omega)$ for the large band gap at different Fermi energies: (a) $V_i=150 \mathrm{mV}, \varepsilon_F=200 \mathrm{meV}$ and (b) $V_i=$ $150 \mathrm{mV}, \varepsilon_F=0 \mathrm{meV}$. The gray lines represent the $\sigma(\omega)$ spectra for the $AB$ domain \hl{at $x=75\mathrm{~nm}$}, and the colored lines show the $\sigma(\omega)$ spectra for solitons \hl{at $x=0\mathrm{~nm}$} with $\varphi=0^{\circ}, 30^{\circ}, 45^{\circ}, 60^{\circ}, 70^{\circ}, 80^{\circ}, 90^{\circ}$. Parameters used for the calculations are $\epsilon=0 \%$, $T=300 \mathrm{~K}$ and $\eta=15 \mathrm{~meV}$.}
\end{figure}

In the following, we systematically investigate the local optical conductivity $\sigma(\omega)$ as a function of soliton angle $\varphi$. We display the $\sigma(\omega)$ spectra at the center of several representative solitons \hl{at $x=0\mathrm{~nm}$} (colored curves) and Bernal stacked region \hl{at $x=75\mathrm{~nm}$} (gray curves) in Fig. ~\ref{fig:LOP1} and Fig.~\ref{fig:LOP2}. These $\sigma(\omega)$ spectra are normalized to monolayer graphene conductivity $\sigma_{\text{mono}}=\frac{e^2}{4 \hbar}$. \hl{We consider three typical cases with different Fermi energies and band gaps for practical graphene devices commonly used in experiments.} In a device with a single electrical gate, the width of bandgap $\Delta$ cannot be independently set from the value of Fermi level $\varepsilon_F$. Hence, based on the experimental work presented in Ref.\cite{RN159}, two sets of realistic values considered in our calculation are $V_i=20 \mathrm{mV}, \varepsilon_F=60 \mathrm{meV}$ (Fig.~\ref{fig:LOP1}) and $V_i=150 \mathrm{mV}, \varepsilon_F=200 \mathrm{meV}$ (Fig.~\ref{fig:LOP2}(a)), respectively. \hl{In the dual-gated bilayer graphene, the band gap $\Delta$ and Fermi energy $\varepsilon_F$ can be individually controlled by top and bottom gate voltages\cite{RN609}, allowing the Fermi level to lie within the bandgap (Fig.~\ref{fig:LOP2}(b)).}

For the case of a narrow gap, the dependence of local optical conductivity spectra $\sigma(\omega)$ on the soliton angle at room temperature is shown in Fig.~\ref{fig:LOP1}(a). A prominent peak in the $\sigma(\omega)$ spectrum at the $\mathrm{AB}(\mathrm{BA})$ region is visible near photon energy $\omega=\gamma_1$. The $\sigma(\omega)$ spectra for solitons are also characterized by a series of resonance peaks, the most notable of which are the sharp peak labeled (i) for the tensile soliton (blue line) and the peak labeled (ii) for the shear soliton (red line). Of particular interest is the distinctive dependence of peak (i) and (ii) on soliton angle from tensile to shear soliton, where peak (i) is gradually suppressed to be comparable to the $\mathrm{AB}(\mathrm{BA})$ domain, whereas peak (ii) emerges and grows into a prominent peak in $\sigma(\omega)$ spectra. Peaks (i) and (ii) both exhibit large enhancement relative to the optical conductivity $\sigma(\omega)$ in $\mathrm{AB}(\mathrm{BA})$ region in the same frequency, which leads to polariton scattering by the solitons and generates strong local optical response in bilayer graphene\cite{RN439}. Instead, the domain wall becomes transparent to the polariton when its optical conductivity $\sigma(\omega)$ is the same as that of the $\mathrm{AB}(\mathrm{BA})$ domain. Therefore, it is expected that the continuous modulation of nanoscale optical response at solitons can be achieved experimentally through the soliton angle $\varphi$ in the mid-infrared range. 

\begin{figure}
\includegraphics[width=1\columnwidth]{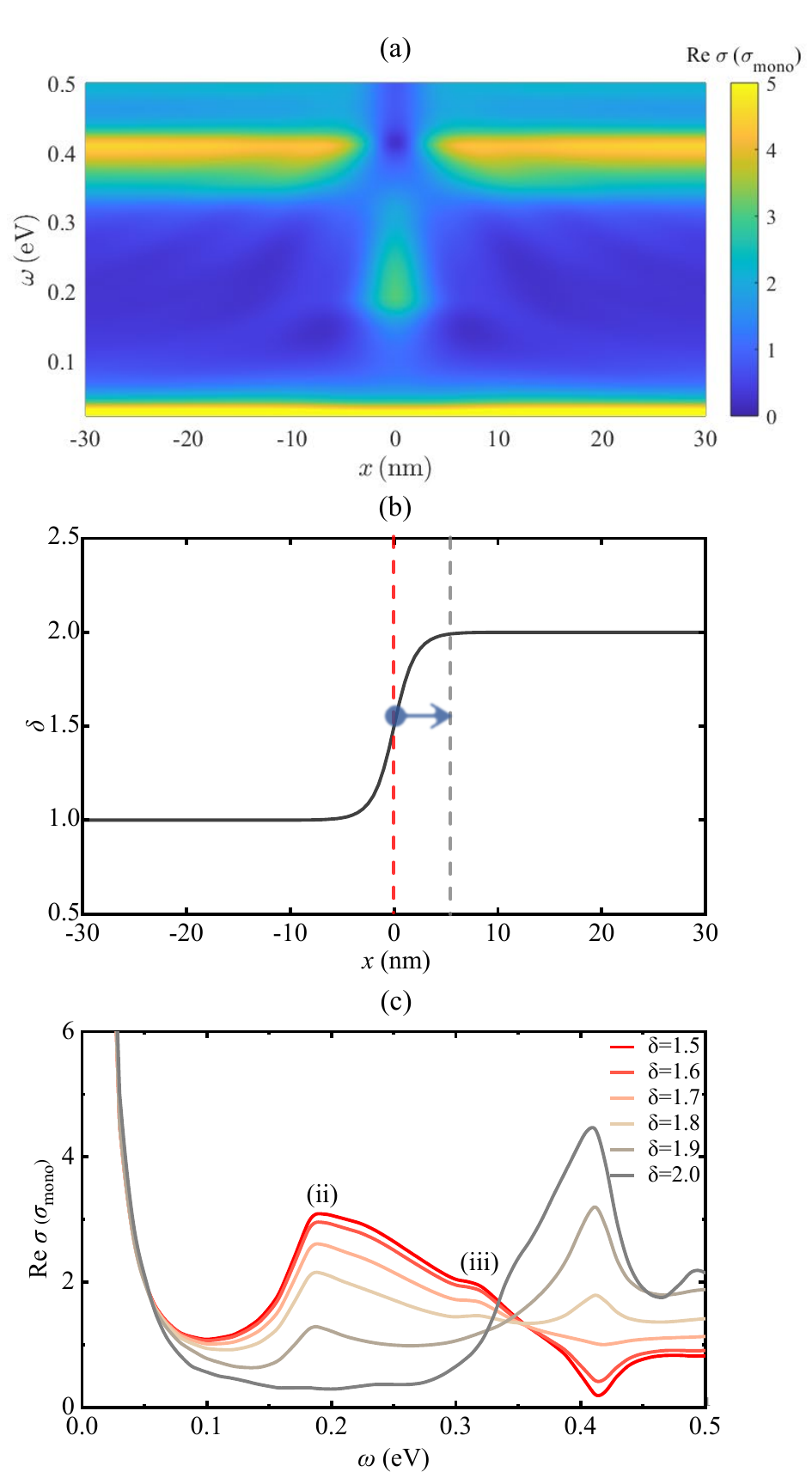}% Here is how to import EPS art
\caption{\label{fig:spatial} (Color online) The spatial dependence of (a) the local optical conductivity spectra $\sigma(\omega)$ and (b) interlayer displacement $\boldsymbol{\delta}$ for shear soliton in bilayer graphene. The center of a soliton is located at $x=0 \mathrm{~nm}$. (c) The $\sigma(\omega)$ collected at equally spaced $\delta$ \hl{(i.e. $x=0,0.5,1.5,2.5,4,10\mathrm{~nm}$)} along the blue arrow direction in (b) corresponds to the transition from solitons (red dashed line) to BA domain (gray dashed line). Parameters used for the calculations are $V_i=150 \mathrm{mV}, \varepsilon_F=200 \mathrm{meV}, \epsilon=0 \%$, $T=300 \mathrm{~K}$ and $\eta=15\mathrm{meV}$.}
\end{figure}

\hl{In order to highlight spectroscopic features in the $\sigma(\omega)$ spectra of solitons, we calculate the local optical conductivity at low temperature ($T=10\mathrm{K}$) with damping parameter $\eta=3\mathrm{meV}$, corresponding to a high electronic mobility sample, as shown in Fig.~\ref{fig:LOP1}(b). The various peaks in the $\sigma(\omega)$ spectra at low temperature become sharper compared with those at room temperature in Fig.~\ref{fig:LOP1}(a), allowing them to be identified clearly. For the AB domain, the $2\varepsilon_F$ behaviour in $\sigma(\omega)$ appears as a visible step feature, which can be attributed to the onset of interband transition at $\omega=2\varepsilon_F$\cite{PhysRevLett.102.037403}. The conductivity below $2\varepsilon_F$ shows nearly zero conductivity up to $2\varepsilon_F$. However, the $2\varepsilon_F$ feature is quite broad in Fig.~\ref{fig:LOP1}(a) due to the thermal broadening. For solitons with soliton angle $\varphi$, the energy of prominent peaks (i) and (ii) can also be well identified in $\sigma(\omega)$ spectra: peak (i) is around $\omega=0.1 \mathrm{eV}$ for the tensile soliton, and peak (ii) is near $\omega=0.23 \mathrm{eV}$ for the shear soliton. Note that the $\sigma(\omega)$ spectra for solitons exhibit a dip at $\omega=\gamma_1$, which will be discussed in Fig.~\ref{fig:spatial}. The main characteristics of all these spectra are similar at different temperatures. Therefore, we will focus on the $\sigma(\omega)$ spectra at room temperature in the following discussion.}

The origin of each resonance peak in the $\sigma(\omega)$ spectrum is discussed below to gain insight into the unique behavior of $\sigma(\omega)$ with respect to soliton angle $\varphi$. For the tensile soliton, peak (i) is attributed to interband transitions between the conventional bound states, which are depicted by a short blue arrow in Fig.~\ref{fig:band}(c). For the shear soliton, the transitions between the $\pm s 1$ states (as well as the topological states) and the high-energy localized states $s 2$, as indicated by the red arrow in Fig.~\ref{fig:band}(a), give rise to the peak (ii) and a shoulder around $\omega=0.3 \mathrm{eV}$ (labeled (iii)) in the $\sigma(\omega)$ spectrum. \hl{The shoulder (iii) involves the optical transition with low spectral weight and thus appears as a weak feature.} Moving from Fig.~\ref{fig:band}(c) to \ref{fig:band}(a), because the conventional bound states are pushed so close to the boundaries of the band continua with increasing soliton angle $\varphi$, the peak (i) in Fig.~\ref{fig:LOP1} is continuously suppressed and redshifts, eventually merging into a Drude peak. The spectral weight loss of peak (i) is balanced by the appearance of the peak (ii) and (iii), due to the transitions associated with the states $s 2$ that gradually emerge as the soliton angle increases. Undoubtedly, the evolution of local band structure profoundly alters the interband transitions involving conventional bound states and high-energy states $(s1$ and $s2)$, resulting in the unique dependence of peak (i) and (ii) on soliton angle in $\sigma(\omega)$ spectra: one peak wanes, the other waxes.

For the large gap case with high $\varepsilon_F$ in Fig.~\ref{fig:LOP2}(a), an examination of Fig.~\ref{fig:LOP1}(a) and \ref{fig:LOP2}(a) shows that the evolution of resonance peak (i) and (ii) in $\sigma(\omega)$ spectra with soliton angle $\varphi$ remains at different bandgap widths $\Delta$. However, the local conductivity at various solitons is greatly enhanced compared to the $\mathrm{AB}(\mathrm{BA})$ domain in the wide frequency range $\omega<0.3 \mathrm{eV}$, even after peak (i) in shear soliton and peak (ii) in tensile soliton are completely suppressed. This feature mainly originates from the suppression of local conductivity spectrum $\sigma(\omega)$ at the $\mathrm{AB}(\mathrm{BA})$ stacking due to the Pauli blocking of interband transitions below $2 \varepsilon_F$. Moreover, the peak (ii) exhibits a larger broadening in Fig.~\ref{fig:LOP2}(a) compared to the narrow gap case in Fig.~\ref{fig:LOP1}(a), accompanied by a shift towards lower energies as the soliton angle $\varphi$ decreases. This behavior can be traced to the imperfect nesting of the dispersion curves for the states $s 2\left(k_{\mathrm{y}}\right)$ and $s 1\left(k_{\mathrm{y}}\right)$ under the large bandgap condition, as well as the density of states $s 2$ redistribution around $E=0.25 \mathrm{eV}$ induced by changes in the soliton angle $\varphi$ as illustrated in Fig.~\ref{fig:band}.

The optical conductivity spectrum at zero Fermi energy provides the possibility of exploring the topological states. We plot the $\sigma(\omega)$ spectra of solitons and $\mathrm{AB}(\mathrm{BA})$ region in Fig.~\ref{fig:LOP2}(b), considering that the Fermi level $\varepsilon_F$ lies within the bandgap due to unintentional doping in the experiment $\left(V_i=150 \mathrm{mV}, \varepsilon_F=0 \mathrm{meV}\right)$. In comparison to the case with the $\varepsilon_F$ above the bulk bandgap in Fig. ~\ref{fig:LOP1}(a) and ~\ref{fig:LOP2}(a), the spectral features of the local conductivity spectra $\sigma(\omega)$ in Fig.~\ref{fig:LOP2}(b) manifest a drastic discrepancy. Here, we begin with a discussion of the interband transitions involved in the $\sigma(\omega)$ spectra, which differs from those in Fig.~\ref{fig:LOP1}(a) and \ref{fig:LOP2}(a). In the uniform $\mathrm{AB}(\mathrm{BA})$ region, the optical transition between the valence and conduction band results in an absorption peak at $\omega=0.15 \mathrm{eV}$, where the energy is equivalent to the bandgap width $\Delta$. \hl{The finite optical response at zero frequency is originated from thermally excited charge carriers at Fermi energy (due to thermal broadening of the Fermi$-$Dirac distribution).} Also, the transition between the valence band and remote conduction band can now occur such that a peak appears at $\omega=\Delta+\gamma_1$ (slightly above $0.5 \mathrm{eV}$). The most intriguing feature is the emergence of an additional peak near $E=0.13 \mathrm{eV}$ in $\sigma(\omega)$ spectra of solitons with $\varphi$ between $60^{\circ}$ and $90^{\circ}$, which is absent in Fig.~\ref{fig:LOP1}(a) and \ref{fig:LOP2}(a), but we still classify it as peak (i) here. The peak (i) undergoes a systematic redshift as the soliton angle $\varphi$ decreases. Notably, in addition to the conventional bound states discussed in Fig.~\ref{fig:LOP1}(a) and \ref{fig:LOP2}(a), the peak (i) in Fig.~\ref{fig:LOP2}(b) is actually the consequence of multiple interband transitions involving topological states and states $\pm s 1$ as well. For solitons with $\varphi$ from $60^{\circ}$ to $90^{\circ}$, peak (i) is dominated by transitions associated with topological states and states $\pm s 1$, due to the conventional bound states are pushed almost to the bulk band boundaries. Additionally, we note that the original peak (ii) vanishes because of the empty $+s 1$ states, while peaks (i) and (iii) have greater spectral weight. Meanwhile, the Drude peak at zero frequency is almost completely depleted since the carrier density in this case is minimized.

Next, we explore the spatial dependence of the local optical conductivity spectra across a domain wall. The spatial distribution of $\sigma(\omega)$ spectra near the soliton is depicted in Fig.~\ref{fig:spatial}(a) for $V_i=150 \mathrm{mV}, \varepsilon_F=200 \mathrm{meV}$, with $\varphi=90^{\circ}$ as an example. Here $x=0 \mathrm{~nm}$ is the center of soliton approximating SP stacking, with uniform AB and BA domains on either side. As seen in Fig.~\ref{fig:spatial}(a), the local optical conductivity at the soliton displays a marked discrepancy compared to that of the $\mathrm{AB}(\mathrm{BA})$ region, indicating a strong dependence on the interlayer displacement $\boldsymbol{\delta}$. The position-dependent stacking order $\delta$ corresponding to various regions in real space is presented in Fig.~\ref{fig:spatial}(b), with red and gray dashed lines denoting the positions of $\delta=1.5$ (SP stacking) and $\delta=2$ (BA stacking), respectively. To provide a step-by-step view of the evolution of electronic properties across the soliton, several $\sigma(\omega)$ spectra at equally spaced stack order $\delta$ variations along the blue arrow direction are plotted in Fig.~\ref{fig:spatial}(c). Since the stacking vector $\boldsymbol{\delta}$ undergoes a transition from SP to uniform BA stacking, the spectral weight of peaks (ii) and (iii) is rapidly depleted, suggesting that the high energy states $(s1$ and $s2)$ are predominantly confined in the soliton's central region. Moreover, \hl{with the transverse from the soliton center at $x=0 \mathrm{~nm}$ to the AB region, the distinct dip at $\omega=\gamma_1$ in $\sigma(\omega)$ for soliton gradually evolves into a prominent peak for AB domain, exhibiting a significant spatial dependence. The feature at $\omega=\gamma_1$ is attributed to interband transitions involving the band at $E=\gamma_1$. However, from the spatial profile of the wave function, the states at $E=\gamma_1$ are only located in the AB (BA) domain and the probability density is almost zero at the soliton, therefore, resulting in a dip near $\omega=\gamma_1$ in $\sigma(\omega)$ for the soliton region.}

\begin{figure}
\includegraphics[width=1\columnwidth]{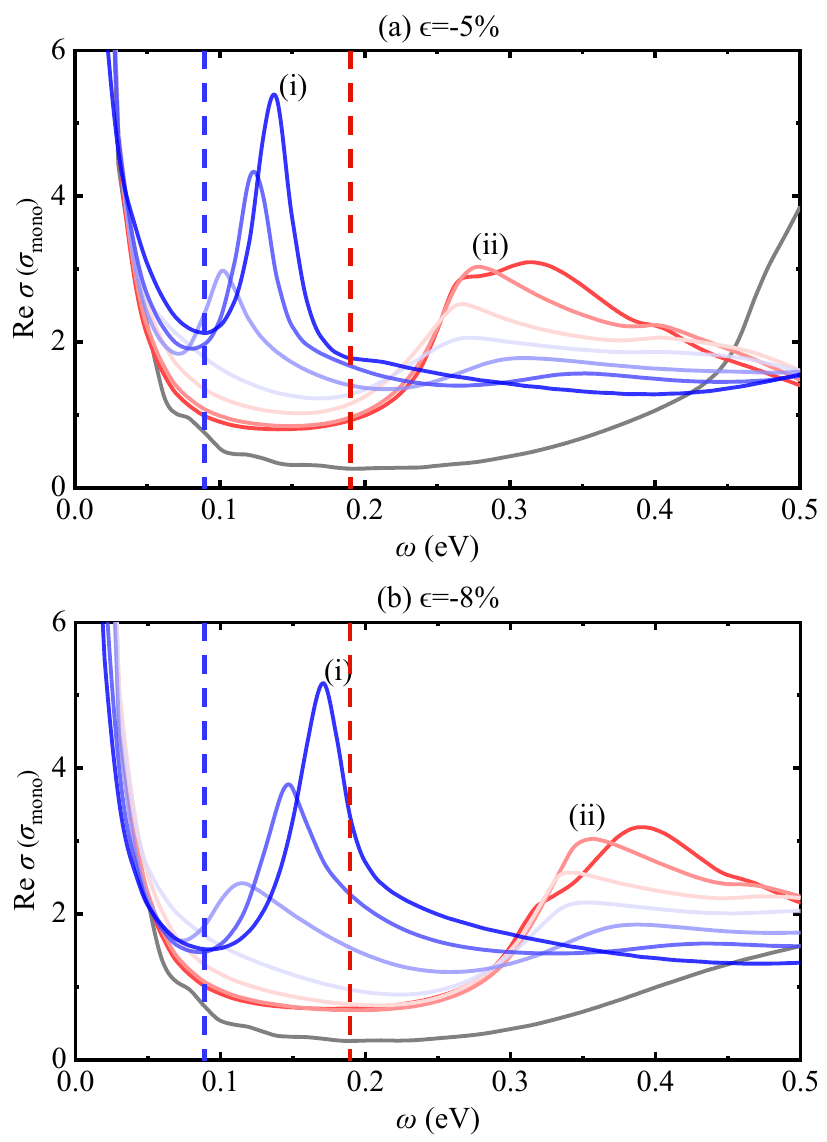}% Here is how to import EPS art
\caption{\label{fig:pressure} (Color online) Real parts of the local optical conductivity spectra $\sigma (\omega)$ under external pressure (a) $\epsilon=-5 \%(P=3.6 \mathrm{GPa})$ and (b) $\epsilon=-8 \%(P= 6.8\mathrm{GPa})$. The vertical blue and red dashed lines represent frequencies corresponding to peaks (i) and (ii) in the $\sigma(\omega)$ without pressure in Fig.~\ref{fig:LOP2}(a). The gray lines represent the $\sigma(\omega)$ spectra for the $AB$ domain \hl{at $x=75\mathrm{~nm}$}, and the colored lines show the $\sigma(\omega)$ spectra for solitons \hl{at $x=0\mathrm{~nm}$} with $\varphi=0^{\circ}, 30^{\circ}, 45^{\circ}, 60^{\circ}, 70^{\circ}, 80^{\circ}, 90^{\circ}$. Parameters used for the calculations are $V_i=150 \mathrm{mV}, \varepsilon_F=200 \mathrm{meV}$, $\eta=15\mathrm{~meV}$ and $T=300 \mathrm{~K}$.}
\end{figure}

Finally, we show that external pressure (commonly used in experiments) can significantly tune the optical properties at the solitons. Fig.~\ref{fig:pressure} shows the local optical conductivity spectra $\sigma(\omega)$ for several values of $\varphi$ under different external pressures $\epsilon$, taking the large bandgap case $\left(V_i=150 \mathrm{mV}, \varepsilon_F=200 \mathrm{meV}\right)$ as an example. The blue and red vertical dashed lines represent frequencies corresponding to peaks (i) and (ii) in the $\sigma(\omega)$ spectra without pressure (Fig.~\ref{fig:LOP2}(a)), respectively. Comparing the case of $\epsilon=-5 \% (P=3.6 G P a)$ in Fig.~\ref{fig:pressure}(a) and $\epsilon=-8 \% (P=6.8 G P a)$ in \ref{fig:pressure}(b), both of which are experimentally achievable pressures, it is clear that the spectral features in $\sigma(\omega)$ evolve dramatically with increasing pressure $\epsilon$, shifting towards higher energies. Especially, the energies of each resonance peak in Fig.~\ref{fig:pressure}(b) are almost twice that of the non-pressure case. Peak (iii) moves to the energies beyond $E=0.5 \mathrm{eV}$ and is therefore not visible within the range displayed in Fig.~\ref{fig:pressure}. In addition, in the $\sigma(\omega)$ spectra of the shear walls, the shoulder slightly above peak (ii) becomes more prominent with the increase of pressure $\epsilon$. The shoulder arises from the interband transitions between the topological states and the $s 2$ band, which can also be observed around $E=0.22 \mathrm{eV}$ in Fig.~\ref{fig:LOP2}(a).

The physics of the remarkable energy doubling of resonance peaks in $\sigma(\omega)$ spectrum with external pressure is now elaborated in detail. To clarify, the significant blue shifts of peaks (i) and (ii) in Fig.~\ref{fig:pressure} are essentially of the same origin, that is, interlayer coupling enhancement due to reduced interlayer spacing caused by pressure. Under the influence of the increased pressure $\epsilon$, the domain walls become sharper as a consequence of lattice reconstruction amplified by enhanced interlayer interaction. For example, the width of tensile soliton narrows from about $10 \mathrm{~nm}$ at no pressure to about $4 \mathrm{~nm}$ at $\epsilon=-8 \%$ as shown in Fig.~\ref{fig:structure}(f). Narrower walls subject the conventional bound states to a stronger quantum confinement effect\cite{RN439}, which results in larger energy separation between the conventional bound states at the same momentum $k_{\mathrm{y}}$ as seen in Fig.~\ref{fig:band}(c) and \ref{fig:band}(f). At the same time, the high-energy soliton state $s 2$ is sensitive to the interlayer interaction $\gamma_1$ related to interlayer spacing. Thus, when vertical pressure $\epsilon$ is applied, the states $s 2$, which was initially near $E=$ $0.25 \mathrm{eV}$, are driven to higher energy by the greater interlayer coupling $\gamma_1$ enhanced by pressure, which exists near $E=0.38 \mathrm{eV}$ with $\gamma_1=0.63 \mathrm{eV}$ at $\epsilon=-8 \%$. Indeed, the external pressure strongly alters the local band structures at solitons such that optical transitions involving these soliton states require higher incident photon energy, which directly contributing to the dramatical blueshift of the associated resonance peak (i), (ii) and (iii) in $\sigma(\omega)$, with energy doubling even at $\epsilon=-8 \%(P=6.8 \mathrm{GPa})$ as shown in Fig.~\ref{fig:pressure}(b).

\section{SUMMARY AND DISCUSSION}
In summary, we investigated systematically the effect of soliton angle as a new and important degree of freedom on the soliton band structure and local optical conductivity in bilayer graphene. Our results suggest that the local electronic structure evolves profoundly with the soliton angle, including the topological states, the high-energy soliton states, and conventional bound states. Many distinctive spectroscopic features appear in the conductivity spectra, which result from interband transitions involving these soliton states. The resonance peaks are highly sensitive to doping, hence the important features in the soliton band structure can be directly measured in experiment. Additionally, we discussed the spatial evolution of local optical conductivity across a domain wall, showing that the soliton states are highly localized in the center of the domain wall. Furthermore, we presented the conductivity spectra for all soliton angles are widely tunable by external pressure, which can double the energies of the resonance peaks at $6.8Gpa$.

Now, we discuss the implications of our work. First, our result reveals that the varied soliton angle very intensely modifies the resonance peaks in local optical conductivity across solitons, with the main peaks ranging from a six-fold increase compared to the AB (BA) region to complete suppression. This result implies that the nanoscale optical response in bilayer graphene can be controlled dramatically through the soliton angle, since the scattering and propagation of polariton near the walls is governed by the local optical conductivity of soliton\cite{RN132,RN439,RN306}. Second, we find that external pressure can greatly tune the resonance-enhanced dynamical conductivity at solitons covering a broad frequency range. This can be used for the modulation of different polaritons in 2D material structures, including plasmon polaritons in graphene\cite{RN132} and phonon polaritons in mTBG/hBN\cite{RN582} as well as mTBG/MoO3 structures\cite{RN400}. Third, we systematically explore the broadband optical conductivity at various solitons for the first time, indicating that the control of nanophotonic properties through soliton angle and pressure can be realized across a broad spectral range. Fourth, our results show the soliton angle can significantly change the local band structure, which is expected to strongly affects the local photo-thermoelectric effect across the solitons and thus can be used to control nanoscale photocurrent generation in bilayer graphene\cite{RN421,RN392}. Finally, because the collective plasmonic mode of the entire soliton network in mTBG is sensitively dependent on the local band structure governed by the soliton angle\cite{RN524}, novel plasmonic effects can be generated in the soliton network through designing and controlling soliton angle profile.

Recent experimental studies have demonstrated the ability to precisely control the soliton networks and single soliton. Previously, the direct manipulation of individual soliton through an atomic force microscope tip\cite{RN594} and soliton anchoring by out-of-plane dislocations\cite{RN592} provide pathways for creating solitons with designed atomic structures. Then, the ability to manipulate the distortion of soliton superlattices is further enhanced by applying the homogeneous uniaxial strain to the substrate via piezoelectric-based mechanical actuation\cite{RN494}. More recently, an innovative approach involving in-plane bending of monolayer ribbons using an atomic force microscope tip has been demonstrated\cite{RN505}, which allows for the controlled creation of soliton superlattices with continuous variation in \hl{moir\'{e}} pattern. With these advances, the soliton angle profile can be potentially designed and controlled in 2D systems. 

\hl{The optical properties of 2D systems, when pressurized in the transparent diamond anvil cells (DACs), can be directly measured by performing in situ photocurrent\cite{Rn607}, photoluminescence\cite{Rn606} and infrared spectroscopy in a wide frequency range\cite{RN515}. With continued recent advances in fabrication techniques, many types of pressure tunable devices can be envisioned. One possible device structure is micron-scale parallel soliton arrays with pre-designed soliton angle featuring desired infrared response. The photoresponse of such soliton arrays can be strongly tuned over a broad spectral range through applying pressure in DACs, which is promising for pressure tunable photonic devices such as photodetectors\cite{RN467}.} Our study reveals the potential to explore a broad spectrum of photonic physics through tuning the soliton angle profile in soliton networks and interlayer coupling in mTBG and related 2D \hl{moir\'{e}} structures.

\section{ACKNOWLEDGMENTS}
This work was supported by the National Natural Science Foundation of China under Grants No. 11874271.

% The \nocite command causes all entries in a bibliography to be printed out
% whether or not they are actually referenced in the text. This is appropriate
% for the sample file to show the different styles of references, but authors
% most likely will not want to use it.

\bibliography{Refs}% Produces the bibliography via BibTeX.
%apsrev4-2.bst 2019-01-14 (MD) hand-edited version of apsrev4-1.bst
%Control: key (0)
%Control: author (8) initials jnrlst
%Control: editor formatted (1) identically to author
%Control: production of article title (0) allowed
%Control: page (0) single
%Control: year (1) truncated
%Control: production of eprint (0) enabled
%apsrev4-2.bst 2019-01-14 (MD) hand-edited version of apsrev4-1.bst
%Control: key (0)
%Control: author (8) initials jnrlst
%Control: editor formatted (1) identically to author
%Control: production of article title (0) allowed
%Control: page (0) single
%Control: year (1) truncated
%Control: production of eprint (0) enabled

%\end{CJK}
\end{document}